\documentclass[pre, 
twocolumn,
amsmath,superscriptaddress, 
floatfix, 
byrevtev]{revtex4-1}
\usepackage{times}
\usepackage{mathptmx}
\usepackage{color}
\usepackage[dvips]{graphicx}

\begin{document}

\title{Two-Dimensional Computation of Pulsed Magnetic Field Diffusion Dynamics in Gold Cone with Consideration of Inductive Heating and Temperature Dependence of Electrical Conductivity}

\author{Hiroki Morita}
\email{morita-h@ile.osaka-u.ac.jp}
\affiliation{Institute of Laser Engineering, Osaka University, 2-6 Yamada-Oka, Suita, Osaka, 565-0871 Japan}
\author{Atsushi Sunahara}
\affiliation{Center of Materials Under eXtreame Environment, Purdue University}
\author{Yasunobu Arikawa}
\affiliation{Institute of Laser Engineering, Osaka University, 2-6 Yamada-Oka, Suita, Osaka, 565-0871 Japan}
\author{Hiroshi Azechi}
\affiliation{Institute of Laser Engineering, Osaka University, 2-6 Yamada-Oka, Suita, Osaka, 565-0871 Japan}
\author{Shinsuke Fujioka}
\email{sfujioka@ile.osaka-u.ac.jp}
\affiliation{Institute of Laser Engineering, Osaka University, 2-6 Yamada-Oka, Suita, Osaka, 565-0871 Japan}

\begin{abstract}
%A high-intesity ($I_\textrm{L} \lambda^2$ $>$ 1.37 $\times$ 10$^{18}$ $\mu$m$^2$ W/cm$^2$) laser pulse produces efficiently a relativistic electron beam (REB) having $>$ 0.511 MeV of the average kinetic energy by interactions with a matter and a plasma.
%The laser-produced REB generally has a large spread angle of $>$ 100 deg. as a typical value, therefore the REB intensity decreases rapidly during long distance transport.
%Reduction of the REB spread is a crucial issue for realizing efficient isochoric heating of a dense matter using the laser-driven REB.
Application of an external kilo-tesla-level magnetic field, which can be generated using high-intensity laser, to a target is a promising scheme to reduce spray angle of a laser-driven relativistic electron beam (REB) for enhancing the isochoric heating of a dense plasma with the laser-driven REB.
%A gold cone is used in some isochoric heating experiments.
%In this scheme, the externally applied magnetic field must diffuse fully in the cone, the REB generation point, and transport region.
%An inductive current is driven by a rapid temporal change of the magnetic field strength in the cone wall, and the inductive current heats ohmically the cone.
%Temperature-dependent electrical conductivity affects significantly the magnetic field diffusion dynamics in the cone wall.
Here we have developed a two-dimensional electro-magnetic dynamics (2D-EMD) simulation code to solve Maxwell equations with considerations of the inductive heating and temperature-dependence of electrical conductivity of a material for calculating temporally and spatially resolved two-dimensional profile of the externally applied magnetic field in a gold-cone-attached target.
\end{abstract}

\maketitle

Strong magnetic fields from a few hundreds to thousand Tesla have been produced by using the capacitor-coil target that is irradiated by high-intensity laser beams \cite{Daido1986, Courtois2005, Fujioka2013, Santos2015, Law2016, Gao2016, Goyon2017} in the laboratories.
This strong magnetic field has already been applied to plasma physics researches, for example, magnetic field reconnection \cite{Pei2016}, control of relativistic electron beam (REB) \cite{Bailly-Grandvaux2018}, magneto-hydrodynamic instability \cite{Matsuo2017} and fast-ignition inertial confinement fusion\cite{Sakata2017}.

Especially in the fast-ignition inertial confinement fusion, relativistic-intensity laser pulse ($I_\textrm{L} \lambda_\textrm{L}^2$ $>$ 1.37 $\times$ $10^{18}$ $\mu$m$^2$ W/cm$^2$) is used to heat instantaneously a pre-compressed fusion fuel, here $I_\textrm{L}$ and $\lambda_\textrm{L}$ are laser intensity and wavelength, respectively.
A gold cone is attached to a spherical fusion fuel for excluding an ablated plasma from the path of the relativistic laser pulse to the pre-compressed fuel.
A high-$Z$ cone is required for preventing the cone wall from breaking by the ablated plasma and preheat caused by radiations and hot electrons.
Relativistic laser-plasma interactions at the cone tip produces efficiently REB having $>$ 0.511 MeV of average kinetic energy.
The laser-produced REB has typically 100 deg. of large divergence angle, and this reduces energy coupling from the heating laser to the fuel core because a small portion of the diverged REB collides with the small fuel core.

Guiding of the diverging REB to the small fuel core is essential for demonstrating efficient fast-ignition.
Application of the strong external magnetic field to the REB generation and transport regions has been proposed by several authors\cite{Strozzi2012, Cai2013, Wang2015, Johzaki2015}, we call this scheme with "Magnetized Fast Ignition (MFI)".
The proof-of-principle experiment of the MFI scheme has been performed at Institute of Laser Engineering Osaka University using the laser-driven capacitor-coil target \cite{Sakata2017}.
The laser-produced magnetic field must diffuse into the REB generation point and REB transport region before producing the REB.

Thickness and open angle of the gold cone used in the experiment \cite{Sakata2017} were 7 $\mu$m and 45 deg., respectively.
The cone tip has a hole whose diameter was 100 $\mu$m diameter.
The 500 $\mu$m-diameter coil made of 50 $\mu$m-diameter nickel wire was located at 225 $\mu$m away the center of fuel.
Magnetic pulse duration is about 1.5 ns of the full width at the half maximum (FWHM) \cite{Law2016}.

Magnetic field diffusion time ($\tau_\textrm{dif}$) is given for a cylindrical material as $\tau_{\rm dif} = 1/2 \mu_0 \sigma r L$, here $\mu_0$, $\sigma$, $r$ and $L$ are permeability, electrical conductivity, cylinder radius and thickness, respectively.
The electrical conductivity of gold is $\sigma$ = 4 $\times$ 10$^7$ S/m at the normal condition.
The magnetic field diffusion time is calculated to be $\tau_\textrm{dif}$ = 8.8 ns for $r$ = 50 $\mu$m and $L$ = 7 $\mu$m, this diffusion time is much longer than the magnetic pulse duration (2 ns) \cite{Law2016}.
The MFI scheme may be spoiled by this slow diffusion of the externally applied magnetic field.

In the real situation, strength of magnetic field produced using the laser-driven capacitor-coil scheme alters rapidly.
This rapid change of the magnetic field strength drives a large inductive current in a cone.
The inductive current heats ohmically the cone.
Electrical conductivity of a material depends strongly on its temperature and relatively weakly on its density.
Temporal change of the electrical conductivity affects significantly the magnetic field diffusion dynamics.

Temperature increment ($\Delta T$) caused by external application of the strong magnetic field pulse can be roughly evaluated as $c_\textrm{V} \rho \Delta T = B_0^2 / 2\mu_0$, here $c_V, \rho, \mu_0$ and $B_0$ are isochoric specific heat, density, permeability and maximum magnetic field strength, respectively, with the assumption that the applied magnetic field energy is completely converted to the material internal energy.
The specific heat of gold at the normal condition is $c_\textrm{V}$ = 129 J/(K kg) and its solid density is $\rho$ = 19.3 g/cm$^3$.
If 500 T magnetic field is applied to a gold cone, its temperature increment is evaluated to be 3.4 eV.
The solid density material at 3.4 eV is in so-called warm dense matter (WDM) state.
Therefore the conductivity of the WDM must be considered to calculate the magnetic field diffusion dynamics.

In this paper, the magnetic field diffusion dynamics in a gold cone was investigated numerically.
We developed a two-dimensional electro-magnetic dynamics (EMD) simulation code to compute spatial- and temporal-resolved two-dimensional magnetic field profiles with considerations of the inductive heating and temperature dependence of electrical conductivities.
A Finite Differential Time Domain (FDTD) method \cite{Schuster2000} was used for solving Maxwell equations.
Inductive heat was calculated with the equation $c_V \rho \partial T/\partial t = \sigma \left( T \right) E^2$, here $E$ is strength of electric field.
The target configuration in this simulation is the same as that in the previous MFI integrated experiment \cite{Sakata2017}.
A temporal shape of a current flow in a coil was a Gaussian, whose FWHM was 1 ns.
The generated magnetic pulse reaches its peak value of 630 T, which corresponds to the maximum current of 250 kA \cite{Law2016}, at 3.0 ns in this simulation.
This calculation runs from 0.0 ns to 6.0 ns.
Computation region and mesh size are to be 5000 $\mu$m $\times$ 5000 $\mu$m and 2 $\mu$m, respectively.

% \subsection{electrical conductivity}

\begin{figure}[bt]
	\centering
	\includegraphics[clip,width=7cm]{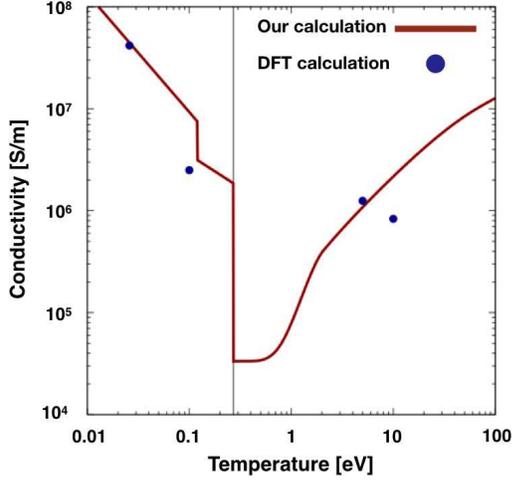}
	\caption{
		Temperature dependence of gold conductivity.
		The red line is the temperature dependence of conductivity in this study.
		The blue points are the conductivities calculated with the DFT.
	}
	\label{fig:cond}
\end{figure}

Figure \ref{fig:cond} shows the temperature dependence of gold conductivity used in this calculation.
The conductivity below  $T$ = 0.27 eV of the boiling point was calculated with the Bloch-Gruneisen formula \cite{Ziman_book}.
The conductivity above the boiling point, which corresponds to WDM state, was calculated with the model \cite{Fu2013} developed by Fu \textit{et al.,}.
The calculated conductivity drops suddenly at a boiling temperature.
This drop is attributed to the metal-insulator transition at the boiling point.
The electrical conductivity of the plasma regime was calculated with modified Spitzer-Harm formula described as
\begin{eqnarray}
	\label{eq:ei_cond}
	% \sigma = \gamma_e\big(Z_{\rm ion}(T)\big) \frac{T^{3/2}}{38 Z_{\rm ion}(T) \ln \big(\Lambda(T)\big)}
	\sigma \left( T \right) = \gamma(Z_{\rm ion})\frac{2^{5/2}}{\pi^{3/2}} \frac{(4 \pi \epsilon_0)^2 (k_B T)^{3/2}}{\sqrt{m_e}e^2 Z_{\rm ion} \ln(\Lambda)}
\end{eqnarray}
where $\gamma$, $\ln(\Lambda)$ and $Z_{\rm ion}$ are the electron correlation factor, Coulomb logarithm and ion charge state respectively.
The electron correlation factor \cite{Bespalov1989} describes the effect of electron-electron collision.
The Coulomb logarithm in this WDM model is calculated with classical-binary cut-off theory\cite{Zaghloul2008}.
And the non-ideal Saha equation, which contains free energy lowering due to the electron screening effect, is used to include the ionization process associated with temperature increment:
\begin{eqnarray}
	\label{eq:saha}
	\frac{n_z n_e}{n_{z-1}} = 2 \frac{U_z}{U_{z-1}}\Big(\frac{2\pi m_e k_B T}{h^2}\Big)^{3/2} \exp{\Big(-\frac{I_z^{\rm eff}}{k_B T}}\Big)
\end{eqnarray}
where $h$, $n_z$ and $U_z$ are Plank constant, number density and internal portion function for $z$-fold ion.
$I_z^{\rm eff}$ is the effective ionization energy from ($z-1$) to $z$.

% \subsection{specific heat}
The specific heat consists of two components, one is caused by ion lattice vibration and the other is by free electrons.
It is known as the Dulong-Petit's law that the ion lattice specific heat is $3 n_\textrm{i} k_B$.
In high temperature above the boiling point, the ion specific heat drops down to $3/2 n_i k_B$ because the lattice structure is broken and the degree of freedom for lattice vibration is vanished.

The specific heat of electrons is obtained by using the chemical potential for a dense plasma, however, the electron specific heat is significantly smaller than the lattice specific heat for low temperature(less than the Fermi temperature $\hbar^2/2m_e (3\pi^2 n_e)^{2/3}$ $\sim$ 1 eV).
When the temperature increases up to a order of 1 eV, the electron specific heat become comparable to lattice one and its value is $3/2 n_e k_B$.
The averaged ionization degree solid gold is calculated as 1.16 at 3 eV from eq.\ref{eq:saha}, majority of gold atoms are singly ionized.
% The gold takes \textcolor{red}{single} ionization state at a several eV because the first ionization energy of gold is \textcolor{red}{10 eV}.
Therefore the specific heat of gold at several eV is approximately estimated as $3/2 (n_\textrm{i} + n_\textrm{e})k_B \sim 3 n_\textrm{i} k_B$.
This value is almost same as the cold specific heat, therefore we used the constant value of specific heat of 129 J/(K kg) in our simulation.

\begin{figure}[bt]
	\centering
	\includegraphics[clip,width=7cm]{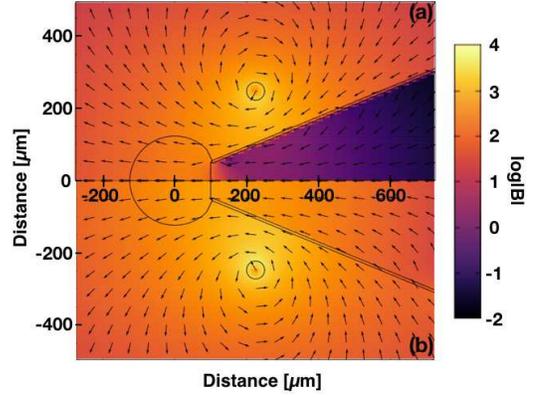}
	\caption{
		Spatial distributions of magnetic field at 3 ns for two cases: (a) the gold conductivity remains the cold value ($\sigma = 4 \times 10^7$S/m) and (b) the temperature dependence of gold conductivity is considered. The applied magnetic field can penetrate through the gold cone wall when the wall temperature increases above the boiling point (0.27 eV).}
	\label{fig:cone_dif}
\end{figure}

Magnetic field diffusion into the gold cone as shown in Fig.\ref{fig:cone_dif} was computed by using 2D-EMD simulation code.
The upper panel of Fig. \ref{fig:cone_dif} shows the spatial profile of magnetic field calculated without consideration of the electrical conductivity change, namely, the gold's electrical conductivity remains its cold value of $4\times10^7$ S/m during this computation.
In this case, the applied magnetic field cannot penetrate the 7 $\mu$m thick cone wall before the peak time (3.0 ns) because the diffusion time evaluated as 8.8 ns as discussed above.
On the other hands, the lower panel in Fig.\ref{fig:cone_dif} shows the profile of magnetic field in the case that the temperature dependence of conductivity shown in Fig.\ref{fig:cond} was considered.
Considering the temperature dependence of conductivity, the applied magnetic field can penetrate through the 7 $\mu$m-thick gold wall.
This is because the gold wall is heated above the boiling point of 0.27 eV at 0.7 ns before the peak timing.

\begin{figure}[bt]
	\centering
	\includegraphics[clip,width=7cm]{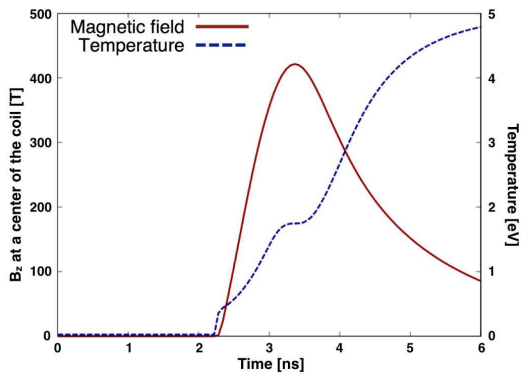}
	\caption{
		Temporal evolutions of magnetic field (red solid line) at the coil center in the cone and temperature (red dash) on the wall outer surface.
		The temperature of the cone wall becomes above the boiling point (0.27 eV) at 2.3 ns, and then the magnetic field penetrates suddenly into the cone.
		Magnetic field reaches maximum value of 421 T at 3.4 ns.
	}
	\label{fig:temp}
\end{figure}

The temporal evolution of magnetic field strength and temperature at ($x$, $y$) = (225, 100)  in the cone wall are shown in Fig.\ref{fig:temp}.
The cone wall temperature was lower than the boiling and magnetic field does not penetrate though the cone wall  before 2.3 ns.
The wall temperature becomes higher than the boiling point at 2.3 ns, then magnetic field begins to diffuse suddenly into the cone.
After that time, the magnetic field rises up and reaches the maximum value at 3.36 ns.
The temperature increase up to 4.79 eV during the pulse duration.

% \section{Discussion}
%The magnetic mirror effect reduces the laser-to-core coupling if $R_\textrm{m}$ is much larger than 20 \cite{Johzaki2017}, here $R_\textrm{m}$ is called as mirror ratio and defined as a ratio between a magnetic field strength at the fuel center and that at the REB generation point, namely the cone tip.

\begin{figure}[bt]
	\centering
	\includegraphics[clip,width=7cm]{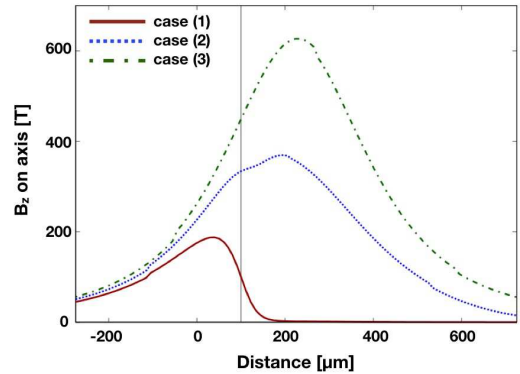}
	\caption{
		Line profiles of the magnetic field strength on axis for three cases. The first case was calculated with the conductivity ($\sigma$ = 1 $\times$ 10$^{-12}$ S/m) of a insulator, the second one is calculated with the conductivity ($\sigma$ = $4 \times 10^7$ S/m) of a cold gold neglecting its temperature dependence, and the third one is calculated with considering the temperature dependence shown in Fig.\ref{fig:cond}.
Magnetic field strengths at the tip of the cone are 96 T, 335 T and 448 T, respectively.}
	\label{fig:cone_comp}
\end{figure}

Our calculation reveals that the temporal change of the electrical conductivity of the cone wall affects significantly magnetic field strength at the cone tip.
Figure \ref{fig:cone_comp} shows line profiles of magnetic field strength along the cone axis calculated for three cases:
The first case is to calculate the profile with the conductivity ($\sigma$ = 1 $\times$ 10$^{-12}$ S/m) of a insulator, the second one is to calculate the profile with the conductivity ($\sigma$ = $4 \times 10^7$ S/m) of a cold gold neglecting its temperature dependence, and the third one is to calculate the profile with considering the temperature dependence shown in Fig. \ref{fig:cond}.
Magnetic fields at the tip of the cone are 96 T, 335 T and 448 T for each case, respectively.

The REB guiding requires $>$ 350 T of the magnetic field strength \cite{Bailly-Grandvaux2018}.
The field strength at the cone tip is not strong enough to guide the REB to core in the case 2.
In the case 3, the field strength at the tip of the gold cone is close to the requirement for the guiding.
Although it was assumed that the capacitor coil target was irradiated by a single beam of GEKKO-XII whose laser energy was 600 J \cite{Law2016}, the capacitor coil target was irradiated by 3 times larger energy in the MFI experiment\cite{Sakata2017}.
The magnetic field at the tip of the cone in Ref.\cite{Sakata2017} could be larger than 350 T that is strong enough to guide the REB.
Our numerical calculation is verified indirectly by the previous experimental result that laser-to-core coupling was improved twice by applying the laser-driven strong magnetic field to the gold-cone-attached target.

In the summary, we have developed the two-dimensional electro-magnetic dynamics simulation code with consideration of the induction heating and temperature dependence of gold conductivity in order to calculate two-dimensional magnetic field profile in the gold cone for the MFI scheme.
The magnetic field penetrates slowly in the gold wall before the timing when its temperature reaches above the boiling point.
Consequently, the peak timing of the interior magnetic field strength is delayed compared to that of exterior one for a 7$\mu$m-thick gold wall.
The calculation shows that the field strength at a tip of gold cone is 335 T that is strong enough guide the REB to fuel core.

%\bibliographystyle{apsrev}
%\bibliographystyle{aip}
%\bibliography{library,book}

\begin{thebibliography}{21}
\expandafter\ifx\csname natexlab\endcsname\relax\def\natexlab#1{#1}\fi
\expandafter\ifx\csname bibnamefont\endcsname\relax
  \def\bibnamefont#1{#1}\fi
\expandafter\ifx\csname bibfnamefont\endcsname\relax
  \def\bibfnamefont#1{#1}\fi
\expandafter\ifx\csname citenamefont\endcsname\relax
  \def\citenamefont#1{#1}\fi
\expandafter\ifx\csname url\endcsname\relax
  \def\url#1{\texttt{#1}}\fi
\expandafter\ifx\csname urlprefix\endcsname\relax\def\urlprefix{URL }\fi
\providecommand{\bibinfo}[2]{#2}
\providecommand{\eprint}[2][]{\url{#2}}

\bibitem[{\citenamefont{Daido et~al.}(1986)\citenamefont{Daido, Miki, Mima,
  Fujita, Sawai, Fujita, Kitagawa, Nakai, and Yamanaka}}]{Daido1986}
\bibinfo{author}{\bibfnamefont{H.}~\bibnamefont{Daido}}
  \bibnamefont{et~al.},
  \bibinfo{journal}{Phys. Rev. Lett.} \textbf{\bibinfo{volume}{56}},
  \bibinfo{pages}{846} (\bibinfo{year}{1986}).

\bibitem[{\citenamefont{Courtois et~al.}(2005)\citenamefont{Courtois, Ash,
  Chambers, Grundy, and Woolsey}}]{Courtois2005}
\bibinfo{author}{\bibfnamefont{C.}~\bibnamefont{Courtois}}
\bibnamefont{et~al.},
\bibinfo{journal}{J. Appl. Phys.}
  \textbf{\bibinfo{volume}{98}}, \bibinfo{pages}{054913}
  (\bibinfo{year}{2005}).

\bibitem[{\citenamefont{Fujioka et~al.}(2013)\citenamefont{Fujioka, Zhang,
  Ishihara, Shigemori, Hironaka, Johzaki, Sunahara, Yamamoto, Nakashima,
  Watanabe et~al.}}]{Fujioka2013}
\bibinfo{author}{\bibfnamefont{S.}~\bibnamefont{Fujioka}}
  \bibnamefont{et~al.}, \bibinfo{journal}{Sci. Rep.}
  \textbf{\bibinfo{volume}{3}}, \bibinfo{pages}{1170} (\bibinfo{year}{2013}).

\bibitem[{\citenamefont{Santos et~al.}(2015)\citenamefont{Santos,
  Bailly-Grandvaux, Giuffrida, Forestier-Colleoni, Fujioka, Zhang, Korneev,
  Bouillaud, Dorard, Batani et~al.}}]{Santos2015}
\bibinfo{author}{\bibfnamefont{J.~J.} \bibnamefont{Santos}}
  \bibnamefont{et~al.}, \bibinfo{journal}{New J. Phys.}
  \textbf{\bibinfo{volume}{17}}, \bibinfo{pages}{083051}
  (\bibinfo{year}{2015}).

\bibitem[{\citenamefont{Law et~al.}(2016)\citenamefont{Law, Bailly-Grandvaux,
  Morace, Sakata, Matsuo, Kojima, Lee, Vaisseau, Arikawa, Yogo
  et~al.}}]{Law2016}
\bibinfo{author}{\bibfnamefont{K.~F.~F.} \bibnamefont{Law}}
\bibnamefont{et~al.},
  \bibinfo{journal}{Appl. Phys. Lett.} \textbf{\bibinfo{volume}{108}},
  \bibinfo{pages}{091104} (\bibinfo{year}{2016}).

\bibitem[{\citenamefont{Gao et~al.}(2016)\citenamefont{Gao, Ji, Fiksel, Fox,
  Evans, and Alfonso}}]{Gao2016}
\bibinfo{author}{\bibfnamefont{L.}~\bibnamefont{Gao}}
  \bibnamefont{et~al.},
  \bibinfo{journal}{Phys. Plasmas} \textbf{\bibinfo{volume}{23}},
  \bibinfo{pages}{043106} (\bibinfo{year}{2016}).

\bibitem[{\citenamefont{Goyon et~al.}(2017)\citenamefont{Goyon, Pollock,
  Turnbull, Hazi, Divol, Farmer, Haberberger, Javedani, Johnson, Kemp
  et~al.}}]{Goyon2017}
\bibinfo{author}{\bibfnamefont{C.}~\bibnamefont{Goyon}}
\bibnamefont{et~al.},
  \bibinfo{journal}{Phys. Rev. E} \textbf{\bibinfo{volume}{95}},
  \bibinfo{pages}{033208} (\bibinfo{year}{2017}).

\bibitem[{\citenamefont{Pei et~al.}(2016)\citenamefont{Pei, Zhong, Sakawa,
  Zhang, Zhang, Wei, Li, Li, Zhu, Sano et~al.}}]{Pei2016}
\bibinfo{author}{\bibfnamefont{X.~X.} \bibnamefont{Pei}}
\bibnamefont{et~al.},
  \bibinfo{journal}{Phys. Plasmas} \textbf{\bibinfo{volume}{23}},
  \bibinfo{pages}{032125} (\bibinfo{year}{2016}).

\bibitem[{\citenamefont{Bailly-Grandvaux
  et~al.}(2018)\citenamefont{Bailly-Grandvaux, Santos, Bellei,
  Forestier-Colleoni, Fujioka, Giuffrida, Honrubia, Batani, Bouillaud, Chevrot
  et~al.}}]{Bailly-Grandvaux2018}
\bibinfo{author}{\bibfnamefont{M.}~\bibnamefont{Bailly-Grandvaux}}
  \bibnamefont{et~al.}, \bibinfo{journal}{Nat. Commun.}
  \textbf{\bibinfo{volume}{9}}, \bibinfo{pages}{102} (\bibinfo{year}{2018}).

\bibitem[{\citenamefont{Matsuo et~al.}(2017)\citenamefont{Matsuo, Nagatomo,
  Zhang, Nicolai, Sano, Sakata, Kojima, Lee, Law, Arikawa et~al.}}]{Matsuo2017}
\bibinfo{author}{\bibfnamefont{K.}~\bibnamefont{Matsuo}}
  \bibnamefont{et~al.}, \bibinfo{journal}{Phys. Rev. E}
  \textbf{\bibinfo{volume}{95}}, \bibinfo{pages}{053204}
  (\bibinfo{year}{2017}).

\bibitem[{\citenamefont{Sakata et~al.}(2017)\citenamefont{Sakata, Lee, Johzaki,
  Sawada, Iwasa, Morita, Matsuo, Law, Yao, Hata et~al.}}]{Sakata2017}
\bibinfo{author}{\bibfnamefont{S.}~\bibnamefont{Sakata}}
\bibnamefont{et~al.},
\bibinfo{journal}{arXiv:1712.06029} (\bibinfo{year}{2017}).

\bibitem[{\citenamefont{Strozzi et~al.}(2012)\citenamefont{Strozzi, Tabak,
  Larson, Divol, Kemp, Bellei, Marinak, and Key}}]{Strozzi2012}
\bibinfo{author}{\bibfnamefont{D.~J.} \bibnamefont{Strozzi}}
  \bibnamefont{et~al.},
  \bibinfo{journal}{Phys. Plasmas} \textbf{\bibinfo{volume}{19}},
  \bibinfo{pages}{072711} (\bibinfo{year}{2012}).

\bibitem[{\citenamefont{Cai et~al.}(2013)\citenamefont{Cai, Zhu, and
  He}}]{Cai2013}
\bibinfo{author}{\bibfnamefont{H.-B.} \bibnamefont{Cai}}
  \bibnamefont{et~al.},
  \bibinfo{journal}{Phys. Plasmas} \textbf{\bibinfo{volume}{20}},
  \bibinfo{pages}{072701} (\bibinfo{year}{2013}).

\bibitem[{\citenamefont{Wang et~al.}(2015)\citenamefont{Wang, Gibbon, Sheng,
  and Li}}]{Wang2015}
\bibinfo{author}{\bibfnamefont{W.-M.} \bibnamefont{Wang}}
  \bibnamefont{et~al.},
  \bibinfo{journal}{Phys. Rev. Lett.} \textbf{\bibinfo{volume}{114}},
  \bibinfo{pages}{015001} (\bibinfo{year}{2015}).

\bibitem[{\citenamefont{Johzaki et~al.}(2015)\citenamefont{Johzaki, Taguchi,
  Sentoku, Sunahara, Nagatomo, Sakagami, Mima, Fujioka, and
  Shiraga}}]{Johzaki2015}
\bibinfo{author}{\bibfnamefont{T.}~\bibnamefont{Johzaki}}
  \bibnamefont{et~al.},
  \bibinfo{journal}{Nucl. Fusion} \textbf{\bibinfo{volume}{55}},
  \bibinfo{pages}{053022} (\bibinfo{year}{2015}).

\bibitem[{\citenamefont{Schuster and Fichtner}(2000)}]{Schuster2000}
\bibinfo{author}{\bibfnamefont{C.}~\bibnamefont{Schuster}} \bibnamefont{and}
  \bibinfo{author}{\bibfnamefont{W.}~\bibnamefont{Fichtner}},
  \bibinfo{journal}{IEEE Trans. Electromagn. Compat.}
  \textbf{\bibinfo{volume}{42}}, \bibinfo{pages}{229} (\bibinfo{year}{2000}).

\bibitem[{\citenamefont{Dharma-Wardana}(2006)}]{Dharma-Wardana2006}
\bibinfo{author}{\bibfnamefont{M.~W.~C.} \bibnamefont{Dharma-Wardana}},
  \bibinfo{journal}{Phys. Rev. E}
  \textbf{\bibinfo{volume}{73}}, \bibinfo{pages}{1} (\bibinfo{year}{2006}).

\bibitem[{\citenamefont{J.M.Ziman}(1972)}]{Ziman_book}
\bibinfo{author}{\bibnamefont{J. M. Ziman}}, \emph{\bibinfo{title}{Principles of
  the Theory of Solids}} (\bibinfo{publisher}{Cambridge University Press},
  \bibinfo{year}{1972}), chap. \bibinfo{chapter}{7 Transport properties}, pp.
  \bibinfo{pages}{223--225}.

\bibitem[{\citenamefont{Fu et~al.}(2013)\citenamefont{Fu, Jia, Sun, and
  Chen}}]{Fu2013}
\bibinfo{author}{\bibfnamefont{Z.}~\bibnamefont{Fu}}
  \bibnamefont{et~al.},
  \bibinfo{journal}{High Energy Density Phys.} \textbf{\bibinfo{volume}{9}},
  \bibinfo{pages}{781} (\bibinfo{year}{2013}).

\bibitem[{\citenamefont{Bespalov and Polishchuk}(1989)}]{Bespalov1989}
\bibinfo{author}{\bibfnamefont{I.~M.} \bibnamefont{Bespalov}} \bibnamefont{and}
  \bibinfo{author}{\bibfnamefont{A.~Y.} \bibnamefont{Polishchuk}},
  \bibinfo{journal}{Sov. Tech. Phys. Lett.} \textbf{\bibinfo{volume}{15}},
  \bibinfo{pages}{39 } (\bibinfo{year}{1989}).

\bibitem[{\citenamefont{Zaghloul}(2008)}]{Zaghloul2008}
\bibinfo{author}{\bibfnamefont{M.~R.} \bibnamefont{Zaghloul}},
  \bibinfo{journal}{Phys. Plasmas} \textbf{\bibinfo{volume}{15}}
  (\bibinfo{year}{2008}).

\end{thebibliography}

\end{document}